\documentclass[letterpaper, 10 pt, conference]{ieeeconf}
\IEEEoverridecommandlockouts  
\title{\LARGE \bf
Coherently coupled quantum oscillators for quantum reservoir computing
}
\usepackage{graphicx}
\usepackage[sorting=none]{biblatex}
\author{Julien Dudas, Julie Grollier and Danijela Markovi\'c$^{*}$
\thanks{Unit\'e Mixte de Physique CNRS, Thales, Universit\'e Paris-Saclay, 91767 Palaiseau, France}
\thanks{$^{*}$Corresponding author: \tt\small danijela.markovic@cnrs.fr}}%

\begin{document}

\maketitle
\thispagestyle{empty}
\pagestyle{empty}

\begin{abstract}

We analyze the properties of a quantum system composed of two coherently coupled quantum oscillators and show through simulations that it fulfills the two properties required for reservoir computing: non-linearity and fading memory. We first show that the basis states of this system apply a set of nonlinear transformations on the input signals and thus can implement neurons. We then show that the system exhibits a fading memory that can be controlled by its dissipation rates. Finally we show that
a strong coupling between the oscillators is important in order to ensure complex dynamics and to populate a number of basis state neurons that is exponential in the number of physical devices.

\end{abstract}

\section{INTRODUCTION}

Artificial neural networks are data based computational models that have revolutionized the field of machine learning over the past ten years. Among their latest achievements are the resolution of a 50 year lasting challenge in biology, i.e. the prediction of protein folding geometry~\cite{Ronneberger2021} and the demonstration of their capacity to assist mathematicians in making fundamental discoveries~\cite{Davies2021}. However, their performance comes at a price of extremely large numbers of neurons, which count in millions for state-of-the-art tasks. Simulating these networks in software leads to long and energy costly training times due to the architecture of  traditional digital computers, in which data is constantly reshuffled between memory and processing units. One way to reduce both time and energy consumption is to implement neural networks directly in hardware, using the approach known as neuromorphic computing~\cite{Merolla2014}. Indeed, neuromorphic computers have distributed architecture, similar to that of brain, with interconnected memory (synapses) and processing (neurons) units, which is more efficient for the implementation of machine learning. However, such physical neural networks are facing the challenge of scaling up, and realization of physical connections between all the neurons on the chip. Different possible solutions to better scaling could by provided by physics~\cite{Markovic2020b, Wright2022}, ranging from the use of dynamics and oscillating neurons, to physical phenomena such as synchronization~\cite{Romera2017a, Romera2022} or energy minimization~\cite{Hopfield}. In particular, possible advantages that quantum physics could bring to artificial neural networks have recently gained a lot of interest~\cite{Schuld2019}. 

Indeed, neural networks can be implemented on quantum hardware~\cite{Biamonte2017}, which has the specificity that its energy levels are discretized, with a quantum basis state corresponding to each energy level. The number of such basis state neurons is exponential in the number of physical components of the system. To realize a quantum neural network, neurons can be encoded in these basis states~\cite{Tacchino2019}, instead of being encoded in individual physical devices. This results in an exponential number of neurons.  The connections between neurons are then implemented as interactions between these basis states, and the number of physical connections that need to be realized is exponentially decreased with respect to classical hardware. 

\begin{figure}
\centering
 \includegraphics[scale=0.5]{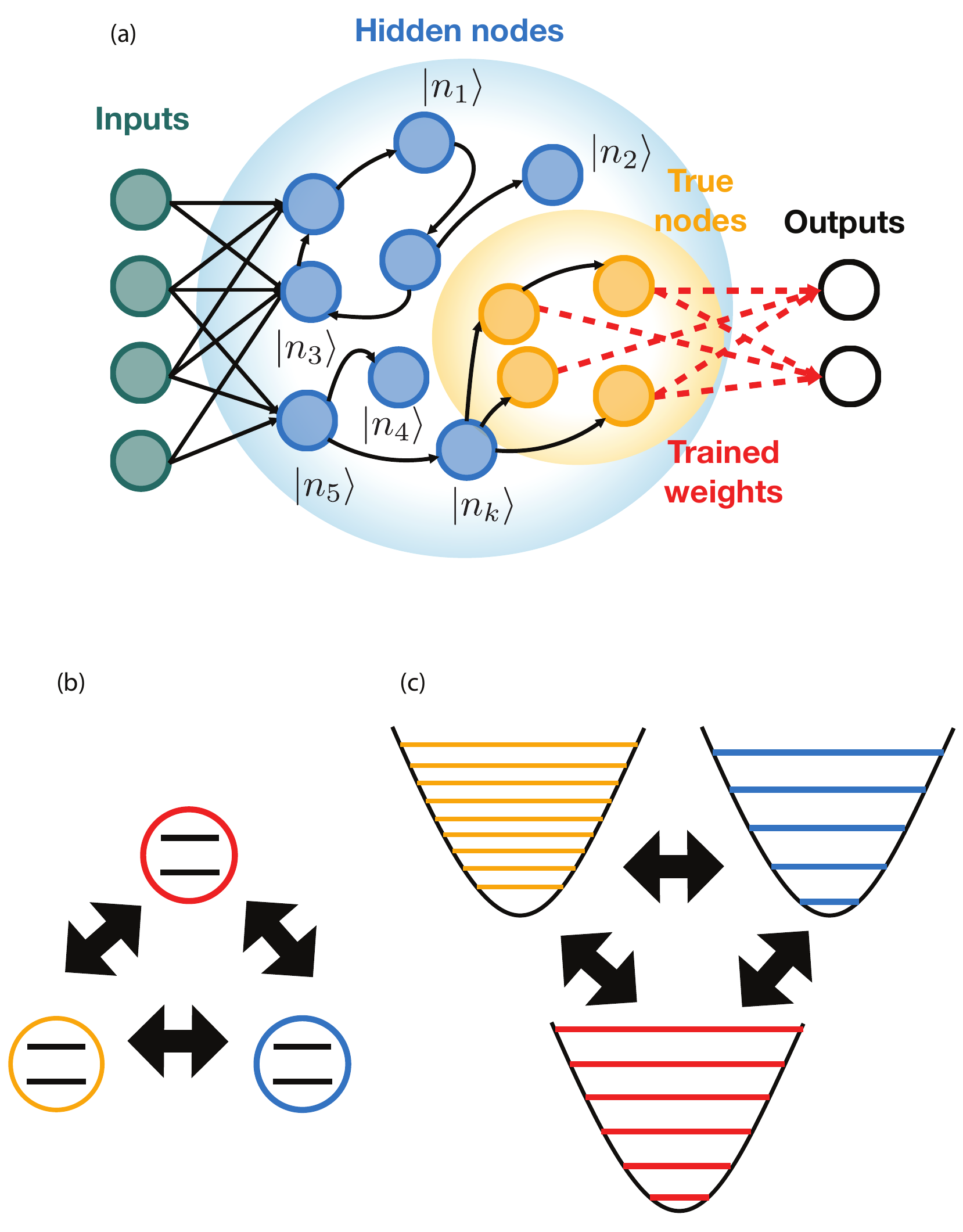}
\caption{(a) Schematic of a quantum reservoir. Input data is represented by green circles. Blue and yellow circles correspond to arbitrarily connected reservoir neurons. Yellow circles are neurons whose values are measured. Black connections are fixed weights and red dashed connections are trained weights. (b) System of coupled qubits. (c) System of coupled quantum oscillators.}
\label{}
\end{figure}

One neural network architecture that allows to exploit a large number of dynamically interacting neurons is reservoir computing. In this article we show that a quantum system composed of coherently coupled quantum oscillators exhibits all the properties needed for the implementation of reservoir computing: a large number of neurons, non-linear and complex neural responses, as well as fading memory.  We first remind the
working principle of reservoir computing, and discuss the requirements a
quantum system needs to satisfy in order to implement
efficient quantum reservoir computing. We then present the model of coupled quantum oscillators that we simulate. 

\begin{figure}
\centering
 \includegraphics[scale=0.7]{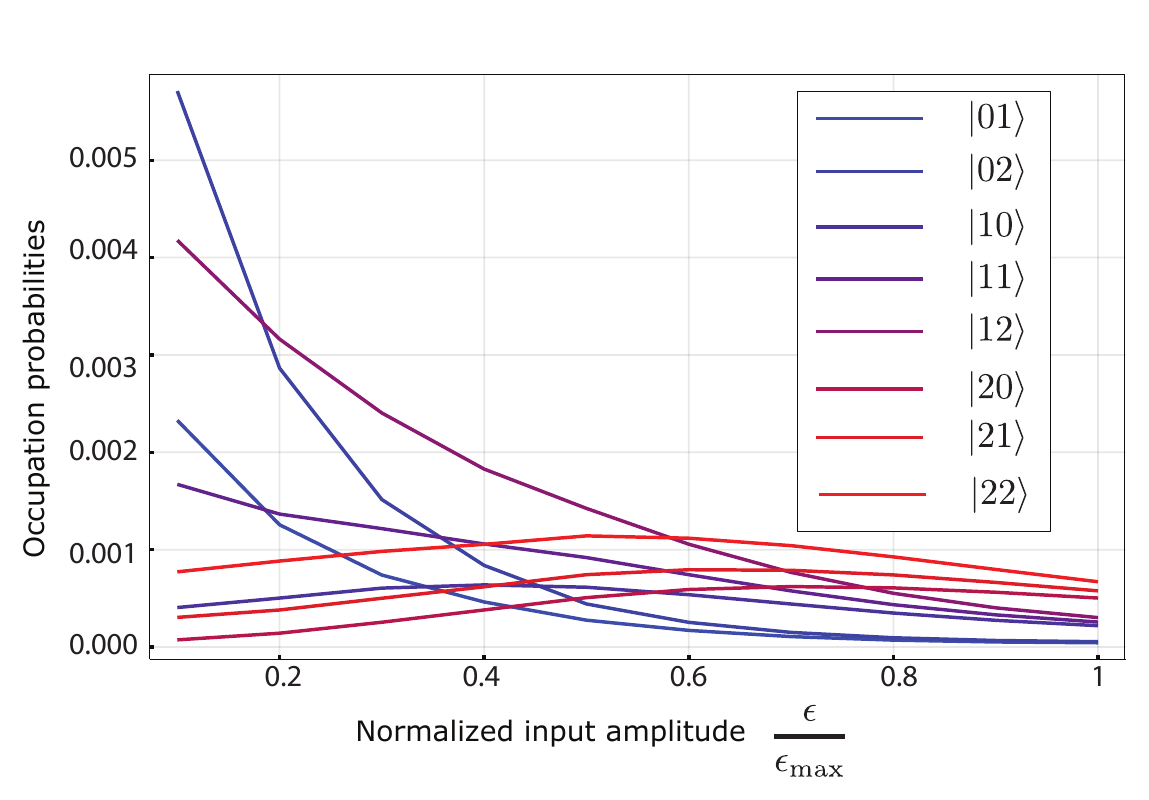}
\caption{Occupation probabilities of levels \{$|01\rangle ....|22\rangle$\} as a function of normalized drive amplitudes. }
\label{nonlinearity}
\end{figure}

\section{RESERVOIR COMPUTING}

A reservoir is a particular type of recurrent neural networks. Recurrent neural networks are a family of neural networks with recurrent connections that provide a memory to the network. This makes them particularly well suited for learning on temporal tasks such as time-series prediction~\cite{Appeltant2011}. The specificity of the reservoir with respect to other recurrent neural networks is that its connections are arbitrary and fixed (black arrows in Figure 1(a)), and only the connections between the reservoir and the output neurons are trained (red dashed arrows in Figure 1(a))~\cite{Haas2004}. The intuition behind reservoir computing is that reservoir neurons perform a complex nonlinear transformation on the input data that projects it into a space where it becomes linearly separable. Linear classification is then done by the fully connected linear layer of weights shown in red dashed arrows Figure 1(a), that can be implemented on a classical digital computer. These weights are trained in a single step by matrix inversion.
The weight matrix $W$ is calculated as
\begin{equation}
    W = {\tilde{Y}_{\rm{train}}} F^\dagger(X_{\rm{train}}),
\end{equation}
where $X_{\rm{train}}$ is the vector containing all the training data, $\tilde{Y}_{\rm{train}}$ is the target vector, and $F^\dagger$ is the Moore-Penrose pseudo-inverse of matrix $F$ containing the outputs of the reservoir neurons for all the training examples. This weights matrix is then applied on the test set data contained in the vector $X_{\rm{test}}$, in order to find the neural network prediction 
\begin{equation}
    Y_{\rm{test}} = W F(X_{\rm{test}}),
\end{equation}
which can be compared to the test target $\tilde{Y}_{\rm{test}}$ in order to evaluate its accuracy.

Reservoir computing was already successfully implemented with different physical systems, such as spintronic nano-oscillators~\cite{RiouIEEE, Torrejon2017, Markovic2019}, optics~\cite{Rafayelyan2020} and electronics~\cite{Appeltant2011}, and learning has been demonstrated on tasks such as spoken digits recognition and multi-dimensional chaotic time-series prediction.

\begin{figure}
\centering
 \includegraphics[scale=0.5]{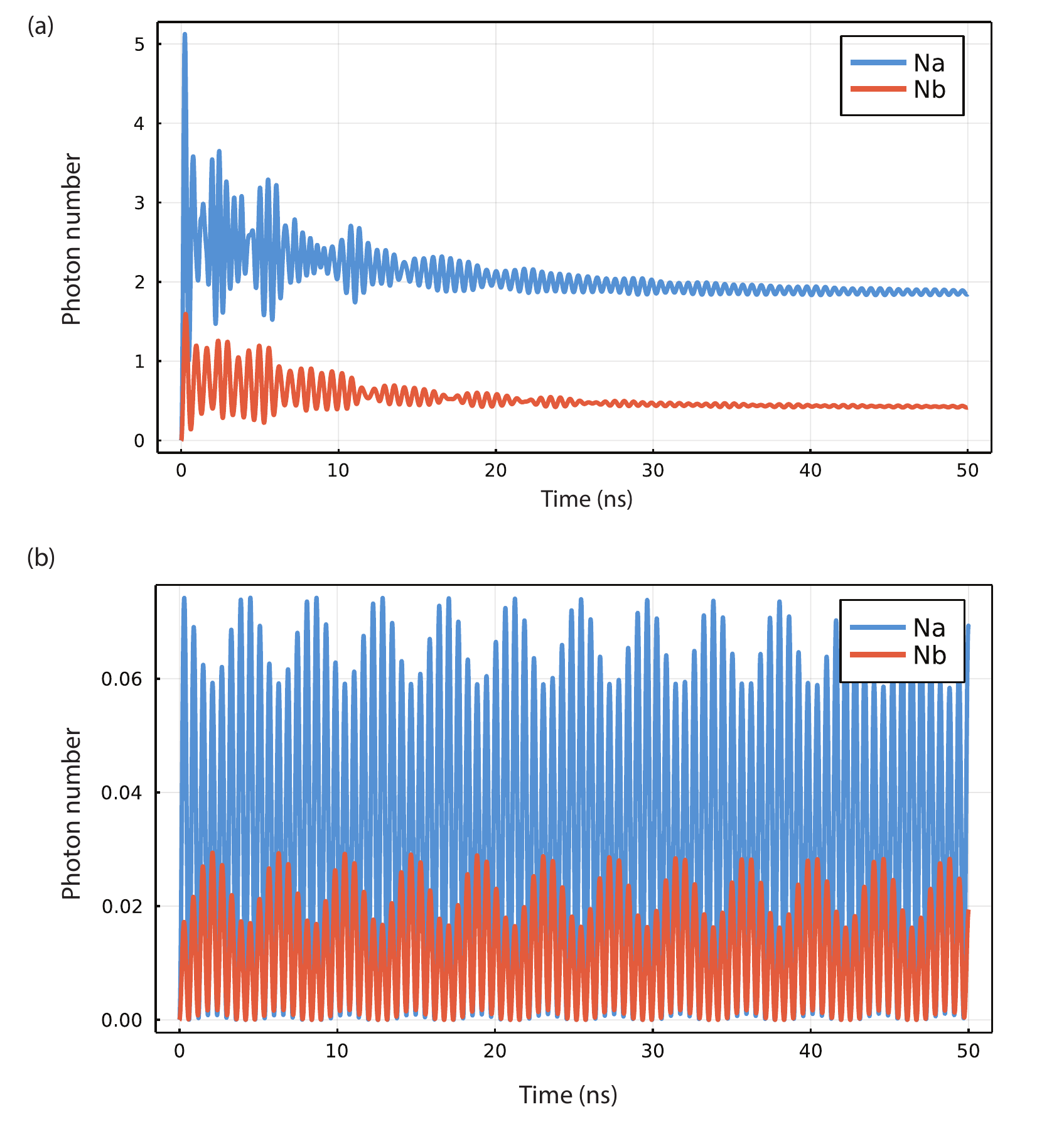}
\caption{Photon number in each resonator for dissipation rates (a) $\kappa_a = \kappa_b = 100$ MHz (b) and for $\kappa_a = \kappa_b = 1$ MHz.}
\label{kappa}
\end{figure}

Quantum reservoir computing was first proposed in 2017 by Fujii and Nakajima~\cite{Fujii2017} with the idea to implement neurons as basis states \{$|00...0\rangle$, $|00...1\rangle$ ....$|11...1\rangle$\} of a system of coupled qubits (Figure 1(b)). The neuron outputs are then obtained by measuring the occupation probabilities of a subset of those basis states (yellow circles in Figure 1 (a)). Quantum reservoir was subsequently simulated on an ensemble of coupled quantum dots, demonstrating its capacity to distinguish entangled and separable input states~\cite{Ghosh2019}, as well as to reconstruct the density matrix of the system providing the input state~\cite{Ghosh2020a}.

 In this work, we consider quantum reservoir implementations on a more general coupled quantum system. In particular, we are interested in coupled multi-level systems such as quantum oscillators (Figure 1(c)). Indeed, a quantum oscillator has an infinite number of discrete energy levels. For weak input signals, we consider that $k$ first energy levels have a finite probability of being populated. For $M$ coupled quantum oscillators we then have a total of $N=k^M$ basis states \{$|n_1\rangle ....|n_N\rangle$\} that can implement neurons. The use of quantum oscillators thus allows to obtain a much better scaling than with qubits. The question we want to answer is whether this system is appropriate to implement a quantum reservoir. Indeed an efficient reservoir has to satisfy two conditions:
\begin{enumerate}
    \item Neurons need to have a nonlinear response to the input drive,
    \item Reservoir needs to have short-term, fading memory.
\end{enumerate}
The first requirement ensures that the input data, once processed by the reservoir, can be separated by a linear classifier. The second requirement is essential for processing temporal sequences for which the output depends not only on the present, but also on the past inputs. Indeed, the reservoir should not remember irrelevant inputs very far in the past, which means that its state should gradually relax to its initial state in the absence of novel inputs.

\section{RESULTS}

To test these properties, we simulate a quantum reservoir with a quantum system composed of two coherently coupled quantum oscillators. The Hamiltonian of this system is
\begin{equation}
    \hat{H}= \hbar \omega_a \hat{a}^\dagger \hat{a} + \hbar \omega_b \hat{b}^\dagger \hat{b} + g\left(\hat{a}{\hat{b}}^\dag+{\hat{a}}^\dag\hat{b}\right), 
\end{equation}
where $\hat{a}$ and $\hat{b}$ are the annihilation operators of the two oscillators, $\omega_a$ and $\omega_b$ are their resonance frequencies and $g$ is their coherent coupling rate. Neurons are then given by the states $|n_a n_b\rangle$ where $n_a$, $n_b$ are the numbers of photons in the oscillators $a$ and $b$. In the simulations we use $\omega_a = 2\pi \times 10$ GHz, $\omega_b = 2\pi \times9.5$ GHz.

We first check the first requirement by verifying that such basis state neurons have a nonlinear response to the input drive. We drive the oscillators at amplitudes $\epsilon_a$ and $\epsilon_b$ that would encode the input data. The drive Hamiltonian writes
\begin{equation}
    \hat{H}_{\rm{drive}} = i \epsilon_a \sqrt{2\kappa_a} (\hat{a} - \hat{a}^\dagger) + i \epsilon_b \sqrt{2\kappa_b} (\hat{b} - \hat{b}^\dagger),
\end{equation}
where $\kappa_a$ and $\kappa_b$ are coupling rates to the transmission lines. We sweep the drive amplitudes in the range
\begin{equation}
\frac{\epsilon}{\epsilon^{\rm{max}}} \in [0,1],
\end{equation}
where $\epsilon_a^{\rm{max}}=10^6 \sqrt{\rm{Hz}}$ and $\epsilon_a^{\rm{max}}=2 \times 10^5 \sqrt{\rm{Hz}}$. We solve the Lindblad master equation
\begin{equation}
    \dot{\rho} = -i [\hat{H} + \dot{H}_{drive}, \rho] + L[C], 
\end{equation}
where the Lindblad superoperator writes
\begin{equation}
    L[C] = \hat{C}\rho \hat{C}^\dagger - \frac{1}{2} \hat{C}^\dagger \hat{C} \rho - \frac{1}{2} \rho \hat{C}^\dagger \hat{C}
\end{equation}
and $\hat{C} = \sqrt{\kappa_a} \hat{a} +  \sqrt{\kappa_b} \hat{b}$ is the collapse operator associated with the decay in the modes $a$ and $b$.
For different drive amplitudes, we measure the occupation probabilities of different basis states $| n_a n_b \rangle$
\begin{equation}
   p(n_a, n_b) = \langle n_a n_b | \rho | n_a n_b \rangle
\end{equation}
 after 100 ns. Occupation probabilities for the basis states $|n_a n_b\rangle \in \{|01\rangle ....|22\rangle$\} as a function of the normalized input drive amplitude are shown in Figure~\ref{nonlinearity}. They show a variety of nonlinear behaviors which proves that they can implement neurons by fulfilling condition 1) stated above. Furthermore, they provide a family of nonlinear functions that can be leveraged by the neural network.

\begin{figure}
\centering
 \includegraphics[scale=0.55]{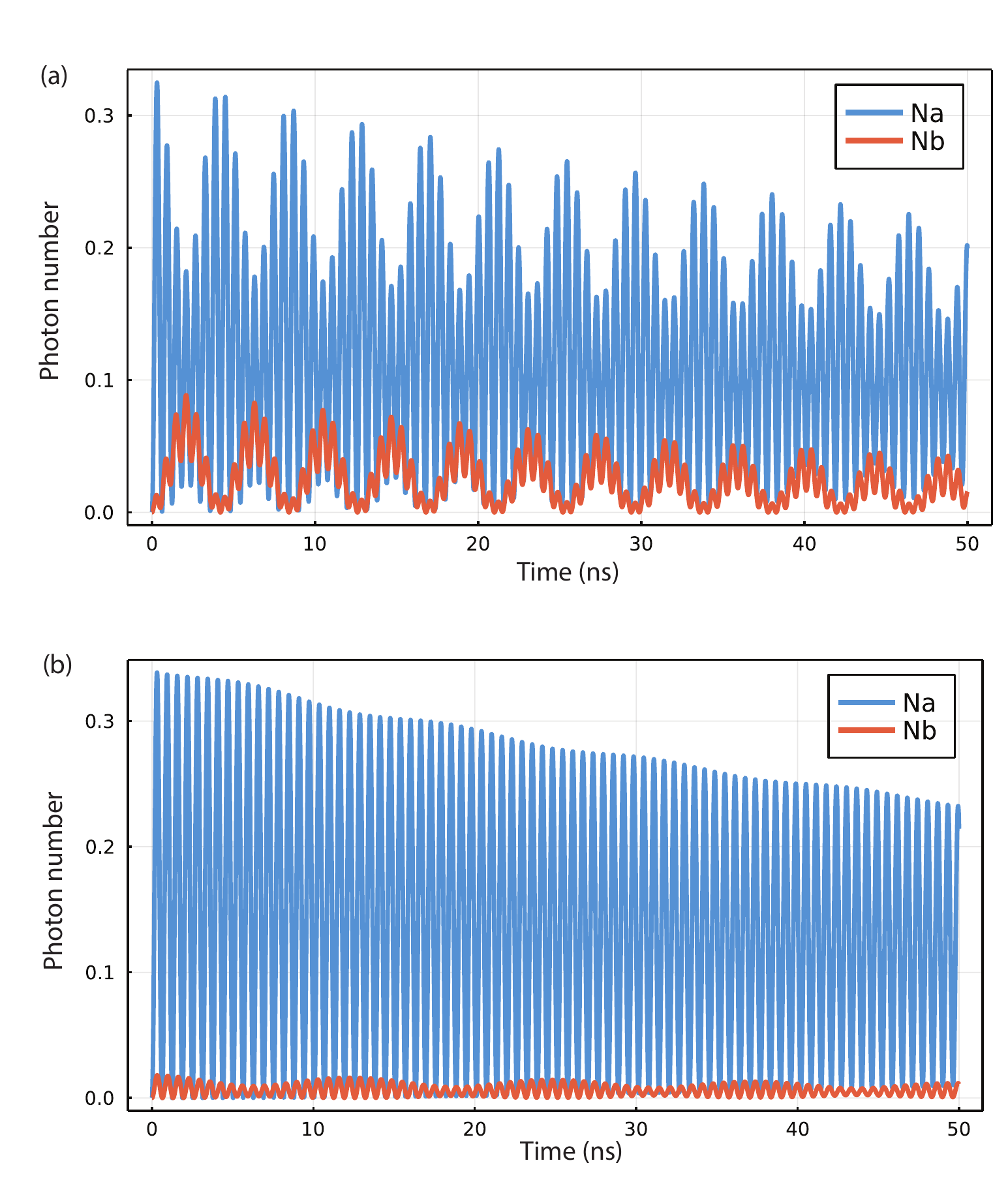}
\caption{Average numbers of photons in each resonator (a) in the strong coupling regime $g$ = 700 MHz and (b) in the weak coupling regime $g$ = 30 MHz.}
\label{coupling}
\end{figure}

In a second time, we investigate the effect of different parameters on the memory of the quantum reservoir. We first fix the coupling rate to $g$ = 700 MHz and the input amplitudes to $\epsilon_a = 10^6 \sqrt{\rm{Hz}}$ and $\epsilon_a = 5 \times 10^5 \sqrt{\rm{Hz}}$, and we look into the effect of the dissipation rates $\kappa_a$ and $\kappa_b$. The measure of the average photon numbers in each oscillator
\begin{equation}
    N_a = \langle \hat{a}^\dagger \hat{a} \rangle
\end{equation}
and
\begin{equation}
    N_b = \langle \hat{b}^\dagger \hat{b} \rangle
\end{equation}
is performed every 2 ns. Their time evolution for a single input is shown in Figure~\ref{kappa} for two different values of dissipation rates. For $\kappa_a = \kappa_b$ = 100 MHz (Figure~\ref{kappa}(a)), the number of photons in the resonator relaxes after 20 ns towards a constant value provided by the drive. Note that number of photons in the oscillator $a$ is larger than in the oscillator $b$ because the drive amplitude is larger. On the contrary, when $\kappa_a = \kappa_b$ = 1 MHz (Figure~\ref{kappa}(b)), we observe a sustained beating in the photon numbers, that comes from the exchange of photons between the two oscillators. For reservoir computing, we want the dissipation rates to be in between these two extrema, such that the reservoir can benefit from the fading memory provided by decoherence and from the complex dynamics provided by coherent coupling.

In the following simulations we fix the dissipation rates to such values between the two extrema, $\kappa_a = 2\pi \times 17$ MHz and $\kappa_b = 2\pi \times 21$ MHz, and we look into the effect of the rate of the coherent coupling between the oscillators. The time evolution of photon numbers for a single input is shown in Figure~\ref{coupling} for two different values of coupling rate $g$. When the coupling rate is much larger than the dissipation rates, the oscillators are in the strong coupling regime (Figure~\ref{coupling}(a)). In this regime, we observe the beating that comes from the exchange of photons between the two oscillators. On the contrary, when the coupling rate is of the same order of magnitude as the dissipation rates, the oscillators are in the weak coupling regime (Figure~\ref{coupling}(b)), where the time evolution is mostly dominated by the dissipation.

\section{DISCUSSION AND CONCLUSION}

In this work, we have simulated coherently coupled quantum oscillators, we have shown that they fulfill all the conditions required for the system to function as a reservoir computer, and we have determined the parameter range in which the system will more favorably compute. In order to benefit from the quantum nature of the reservoir, and its state space exponential in the number of physical components, one has to populate a large number of basis states. Our results highlight that this can only be obtained if the sub-systems are strongly coupled and if dissipation is sufficiently weak. Nevertheless, dissipation is important to a certain degree for reservoir computing because it provides the fading memory needed to classify times series inputs. The ideal parameter set therefore constitutes a compromise between these different features.  This work represents an important step for the future implementation of quantum
reservoirs with coupled oscillators.

\section{Acknowledgements}

This research was supported by the Quantum Materials for Energy Efficient Neuromorphic Computing (Q-MEEN-C), an Energy Frontier Research Center funded by the U.S. Department of Energy (DOE), Office of Science, Basic Energy Sciences (BES), under Award DE-SC0019273.

\addtolength{\textheight}{-12cm}   

\printbibliography

@article{Ronneberger2021,
author = {Ronneberger, Olaf and Tunyasuvunakool, Kathryn and Bates, Russ and {\v{Z}}{\'{i}}dek, Augustin and Ballard, Andrew J and Cowie, Andrew and Romera-paredes, Bernardino and Nikolov, Stanislav and Jain, Rishub and Adler, Jonas and Back, Trevor and Petersen, Stig and Reiman, David and Clancy, Ellen and Zielinski, Michal and Steinegger, Martin and Pacholska, Michalina and Berghammer, Tamas and Bodenstein, Sebastian and Silver, David and Vinyals, Oriol and Senior, Andrew W and Kavukcuoglu, Koray},
journal = {Nature},
number = {May},
pages = {583--589},
publisher = {Springer US},
title = {{Highly accurate protein structure prediction with AlphaFold}},
volume = {596},
year = {2021}
}

@article{Davies2021,
author = {Davies, Alex and Veli{\v{c}}kovi{\'{c}}, Petar and Buesing, Lars and Blackwell, Sam and Zheng, Daniel and Toma{\v{s}}ev, Nenad and Tanburn, Richard and Battaglia, Peter and Blundell, Charles and Juh{\'{a}}sz, Andr{\'{a}}s and Lackenby, Marc and Williamson, Geordie and Hassabis, Demis and Kohli, Pushmeet},
journal = {Nature},
pages = {70--74},
pmid = {34853458},
title = {{Advancing mathematics by guiding human intuition with AI}},
volume = {600},
year = {2021}
}

@article{Biamonte2017,
author = {Biamonte, Jacob and Wittek, Peter and Pancotti, Nicola and Rebentrost, Patrick and Wiebe, Nathan and Lloyd, Seth},
journal = {Nature},
pages = {195--202},
publisher = {Nature Publishing Group},
title = {{Quantum machine learning}},
volume = {549},
year = {2017}
}

@article{Tacchino2019,
author = {Tacchino, Francesco and Macchiavello, Chiara and Gerace, Dario and Bajoni, Daniele},
journal = {npj Quantum Information},
pages = {1--8},
publisher = {Springer US},
title = {{An artificial neuron implemented on an actual quantum processor}},
volume = {5},
year = {2019}
}

@article{Fujii2017,
author = {Fujii, Keisuke and Nakajima, Kohei},
journal = {Physical Review Applied},
pages = {024030},
title = {{Harnessing disordered-ensemble quantum dynamics for machine learning}},
volume = {8},
year = {2017}
}

@article{Haas2004,
author = {Haas, Harald and Jaeger, Herbert},
journal = {Science},
pages = {78--80},
title = {{Harnessing Nonlinearity: Predicting Chaotic Systems and Saving Energy in Wireless Communication}},
volume = {304},
year = {2004}
}

@article{Merolla2014,
author = {Merolla, Paul A and Arthur, John V and Alvarez-icaza, Rodrigo and Cassidy, Andrew S and Sawada, Jun and Akopyan, Filipp and Jackson, Bryan L and Imam, Nabil and Guo, Chen and Nakamura, Yutaka and Brezzo, Bernard and Vo, Ivan and Esser, Steven K and Appuswamy, Rathinakumar and Taba, Brian and Amir, Arnon and Flickner, Myron D and Risk, William P and Manohar, Rajit and Modha, Dharmendra S},
journal = {Science},
pages = {6197},
title = {{A million spiking-neuron integrated circuit with a scalable communication network and interface}},
volume = {345},
year = {2014}
}

@article{Appeltant2011,
author = {Appeltant, L. and Soriano, M. C. and {Van Der Sande}, G. and Danckaert, J. and Massar, S. and Dambre, J. and Schrauwen, B. and Mirasso, C. R. and Fischer, I.},
journal = {Nature Communications},
pages = {468},
pmid = {21915110},
publisher = {Nature Publishing Group},
title = {{Information processing using a single dynamical node as complex system}},
volume = {2},
year = {2011}
}

@article{Torrejon2017,
author = {Torrejon, Jacob and Riou, Mathieu and Araujo, Flavio Abreu and Tsunegi, Sumito and Khalsa, Guru and Querlioz, Damien and Bortolotti, Paolo and Cros, Vincent and Yakushiji, Kay and Fukushima, Akio and Kubota, Hitoshi and Yuasa, Shinji and Stiles, Mark D and Grollier, Julie},
journal = {Nature},
pages = {428--431},
publisher = {Nature Publishing Group},
title = {{Neuromorphic computing with nanoscale spintronic oscillators}},
volume = {547},
year = {2017}
}

@article{RiouIEEE,
author = {Riou, M. and Araujo, F. Abreu and Torrejon, J. and Tsunegi, S. and Khalsa, G. and Querlioz, D. and Bortolotti, P. and Cros, V. and Yakushiji, K. and Fukushima, A. and Kubota, H. and Yuasa, S. and Stiles, M. D. and Grollier, J.},
journal = {IEEE Trans Electron Devices},
title = {{Neuromorphic Computing through Time-Multiplexing with a Spin-Torque Nano-Oscillator}},
year = {2017}
}

@article{Markovic2019,
author = {Markovi{\'{c}}, Danijela and Leroux, N and Riou, M and {Abreu Araujo}, F and Torrejon, J and Querlioz, D and Fukushima, A and Yasa, S and Trastoy, J and Bortolotti, P and Grollier, J},
journal = {Applied Physics Letters},
pages = {012409},
title = {{Reservoir computing with the frequency, phase, and amplitude of spin-torque nano- oscillators}},
volume = {114},
year = {2019}
}

@article{Rafayelyan2020,
author = {Rafayelyan, Mushegh and Dong, Jonathan and Tan, Yongqi and Krzakala, Florent and Gigan, Sylvain},
journal = {Physical Review X},
pages = {41037},
publisher = {American Physical Society},
title = {{Large-Scale Optical Reservoir Computing for Spatiotemporal Chaotic Systems Prediction}},
volume = {10},
year = {2020}
}

@article{Ghosh2019,
author = {Ghosh, Sanjib and Opala, Andrzej and Matuszewski, Micha{\l} and Paterek, Tomasz and Liew, Timothy C. H.},
journal = {npj Quantum Information},
title = {{Quantum reservoir processing}},
volume = {5},
year = {2019}
}

@article{Ghosh2020a,
author = {Ghosh, Sanjib and Opala, Andrzej and Matuszewski, Micha{\l} and Paterek, Tomasz and Liew, Timothy C.H.},
journal = {IEEE Transactions on Neural Networks and Learning Systems},
pages = {1--18},
title = {{Reconstructing quantum states with quantum reservoir networks}},
year = {2020}
}

@article{Markovic2020b,
author = {Markovi{\'{c}}, Danijela and Mizrahi, Alice and Querlioz, Damien and Grollier, Julie},
journal = {Nature Reviews Physics},
pages = {499--510},
title = {{Physics for neuromorphic computing}},
volume = {2},
year = {2020}
}

@article{Romera2017a,
author = {Romera, Miguel and Talatchian, Philippe and Tsunegi, Sumito and Araujo, Flavio Abreu and Cros, Vincent and Bortolotti, Paolo and Yakushiji, Kay and Fukushima, Akio and Kubota, Hitoshi and Yuasa, Shinji and Vodenicarevic, Damir and Locatelli, Nicolas and Querlioz, Damien and Grollier, Julie},
journal = {Nature},
pages = {230--234},
pmid = {30374193},
publisher = {Springer US},
title = {{Vowel recognition with four coupled spin-torque nano-oscillators}},
volume = {563},
year = {2018}
}

@article{Hopfield,
author={Hopfield, J J.},
title={Neural networks and physical systems with emergent collective computational abilities},
journal={Proceedings of the National Academy of Sciences of the United States of America},
volume={79(8)},
year={1982},
pages={2554–2558}
}

@article{Schuld2019,
author = {Schuld, Maria and Killoran, Nathan},
journal = {Physical Review Letters},
pages = {40504},
publisher = {American Physical Society},
title = {{Quantum Machine Learning in Feature Hilbert Spaces}},
volume = {122},
year = {2019}
}

@article{Romera2022,
author = {Romera, Miguel and Talatchian, Philippe and Tsunegi, Sumito and Yakushiji, Kay and Fukushima, Akio and Kubota, Hitoshi and Yuasa, Shinji and Cros, Vincent and Bortolotti, Paolo and Ernoult, Maxence and Querlioz, Damien and Grollier, Julie},
journal = {Nature Communications},
number = {1},
pages = {1--7},
pmid = {35169115},
publisher = {Springer US},
title = {{Binding events through the mutual synchronization of spintronic nano-neurons}},
volume = {13},
year = {2022}
}

@article{Wright2022,
author = {Wright, Logan G and Onodera, Tatsuhiro and Stein, Martin M and Wang, Tianyu and Schachter, Darren T and Hu, Zoey and Mcmahon, Peter L},
journal = {Nature},
pages = {549--555},
publisher = {Springer US},
title = {{Deep physical neural networks trained with backpropagation}},
volume = {601},
year = {2022}
}

\end{document}